# Furnace for Inelastic X-Ray Scattering from Liquids to 1600C

*Alfred Q.R. Baron[1]\*, Masanori Inui[2], Daisuke Ishikawa[1], Kazuhiro Matsuda[3], Yukio Kajihara[2], Yoichi Nakajima[4], Kazuhiko Taguchi[1] and Yasunori Hattori[1]*

[1]Materials Dynamics Laboratory, RIKEN SPring-8 Center, 1-1-1 Kouto, Sayo, Hyogo 679-5148, Japan
[2]Graduate School of Advanced Science and Engineering, Hiroshima University, Higashi-Hiroshima, 739-8521, Japan
[3]Graduate School of Science, Kyoto University, Kyoto 606-8502, Japan
[4]Department of Physics, Kumamoto University, Kumamoto 860-8555, Japan

*baron@spring8.or.jp

Keywords: furnace, inelastic X-ray scattering, liquid dynamics, dynamic structure factor

The design and implementation of a furnace for inelastic x-ray scattering from liquids with sample temperatures up to ~1600C is described. Carbon composite heaters operating in vacuum provide robust heating elements: one pair of heaters has been used for >18 days of operational time above 1500C, including 8 cycles to room temperature. High quality data has been obtained to scattering angles as low as 7 mrad in two-theta (Q<1 nm$^{-1}$ at 25.7 keV) from a sample at 1560C.

## 1. Introduction

The behavior of liquids on atomic length scales is an area of current scientific investigation. In particular, with the advent of practical inelastic x-ray scattering (IXS) setups at modern synchrotron radiation facilities which are not subject to the kinematic limitations of neutron scattering, experimental investigation of angstrom- and nm-scale dynamics of liquids has greatly expanded (see, e.g., discussion in [1]). However, many materials melt only at relatively high temperature (relevant references on liquid iron, which melts at 1538C, and will be shown as an example below, include [2–6]) so, to extend the range of samples investigated, an IXS-compatible furnace is needed. Furthermore, this furnace must be stable on ~day time scales (or longer) as, even at modern x-ray sources, count rates in inelastic x-ray scattering experiments are often low, so that spectra may be measured for ~24 hours. Finally, as it is highly desirable to extend IXS investigations to small scattering angles (long length scales), where signal rates can be very low, the background from the heating system (windows, cell materials) must be low, especially at small angles. Such a system, in principle, may also be interesting for x-ray small- and wide-angle scattering measurements to investigate, e.g. isothermal compressibility, or liquid-liquid phase changes, or other topics of current interest in disordered materials at high temperature.

Here we describe a furnace optimized for use at the high (meV-)-resolution IXS spectrometer of BL43LXU [7,8] of the RIKEN SPring-8 Center. The furnace has been operating for about 2 years over >6 experimental runs. We find it to be reliable and straightforward to operate up to ~1600C. This furnace was made after difficulties with another system [9] made it impossible to collect comprehensive IXS data sets: the heaters tended to fail on ~hour time scales for higher temperatures (T>~1300C). The older system had performed better previously, e.g., [2], but did not recover despite multiple weeks of work. The new system has three notable changes compared to that in [9]: (1) the use of carbon composite heating elements which are robust and allow repeated operation at high (~1600C) temperature for multi-day runs, (2) a design that uses contact cooling initiated by thermal expansion and (3) operation in vacuum. Additional details are discussed below, including the over-all concept, material and heating element choices, practical details for using thermocouples, and the performance.

**2. Container vs Containerless Sample Support**

An important distinction for high-temperature work is between a containerless sample environment and a container based one. Containerless setups are possible using levitation via gas flow or electrostatic methods. They have the advantage that the sample is intrinsically thermally isolated from nearby solid materials, so that one does not have to worry about, e.g., exceeding the melting point of the surrounding material, thus allowing extreme (>3000C) temperatures with laser heating. Containerless setups also can facilitate supercooling of liquids. However, levitation usually requires mm-scale (e.g., ~2 mm diameter) samples for the control/feedback on the sample position. This means that while such methods work well for relatively low-Z/light materials such as alumina [10] where x-ray attenuation lengths are ~mm so the experiment is done in transmission, they have not been used for high-Z/heavy materials, where the attenuation length of the ~20 keV x-rays used for most IXS work is on the 0.01 to 0.1 mm scale. There has been at least one case where levitation was used with a reflection geometry (liquid Ti [11], attenuation length ~0.2 mm). If the position control is sufficiently good, such a geometry might also be used for heavier materials with shorter attenuation lengths, though, even then, measurements at small scattering angles may be tricky (the work presented in [11] was entirely above 5 nm$^{-1}$). However, in our experience [12], , the position of levitated samples can fluctuate at the level of a few 0.01, so a reflection geometry may be difficult. Further, as a levitated sample is not in a sealed environment, there can be issues with sample evaporation (e.g. the lifetime of a 2 mm diameter sphere of liquid silicon electrostatically levitated in vacuum was only a few hours).

Container based methods allow flexibility to tailor the sample thickness down to the level of 0.1 or even, with care, to ~0.01 mm. Further, the containers may be sealed so that sample evaporation is not an issue. In addition, given an appropriate chamber, they can be operated at relatively high pressures (~kbar) as is interesting to explore the liquid phase diagram (see, e.g., [13]), though high pressure is not the goal in the present work. However, the melting point of the container material will limit the maximum achievable temperature. Also, scattering from the container can create background (as is also an issue for gas flow levitation), so that background subtraction is often needed, and the container material must be chosen with care to avoid the background dominating the signal. The background can be particularly problematic when data at small scattering angles is desired. Single crystal sample containers, or at least single crystal windows, are often preferable as they have less small angle scattering than other (polycrystalline or glassy) materials, and will have well-defined phonon modes that can often be subtracted from data relatively easily.

The present work relies on the single crystal sapphire cell technology developed by Tamura, Inui and Hosokawa at Hiroshima University [14]. The single crystal material means that the small angle scattering background from the container can be reduced, while the relative hardness of the sapphire means the inelastic (phonon) background from the cell will tend, at least for small momentum transfers, to be at high energy transfers so is often out of the energy window of interest for liquids. The long shape of these sapphire cells (see [14]) with the sample position about 25mm from where the cell is sealed allows the sample temperature to be different than the temperature of the seal. This means that (1) it is possible to seal the cell

(melting a glass at ~1200C) without heating the sample too much and (2) in use the sample can be heated to high temperature in the interior of the furnace, while the sealing, far from the center of the furnace, remains relatively cool and does not melt, preserving the integrity of the cell. While the use of the sapphire does limit the maximum temperature (the melting point of sapphire is ~2040C) there are still many different materials that melt in the nominally accessible temperature range (say, < 1900C). The present furnace is designed for use with Hiroshima style sapphire cells.

## 3. Materials

Material choices are critical for proper operation of the system. Table 1 details some of the materials used, their purpose, and approximate limits on upper temperature.

| Material | Melting Point ($T_M$) or Maximum Use Temperature (C) | Comment |
| --- | --- | --- |
| Copper | $T_M$ = 1084 | Soft – used primarily for electrical connectors. (beryllium copper preferred for structural parts) |
| Beryllium Copper | $T_M$ ~ 850-905 (depending on alloy) | Structural alternative to copper |
| Aluminum | $T_M$ ~ 560-660 (depending on alloy) | Easy to machine, light weight, good thermal conductor |
| Molybdenum | $T_M$ = 2623 | Machinable with effort, sheets (0.1 or 0.05 mm thick) readily available. |
| Tungsten | $T_M$ = 3422 | Difficult to Machine |
| Stainless Steel | $T_M$ ~1400-1500 (depending on alloy) | Relatively poor thermal conductivity. |
| Alumina | $T_M$ ~2070 | Hard, Thermal Insulator |
| Sapphire | $T_M$ = 2040 | Hard, Sample Cell |
| Macor | ~1000 | Insulator (machinable ceramic used as electrical insulator) |
| Hexagonal BN | ~3000 (sublimation) | Soft (can be cut with a knife) but slow delivery and expensive. |
| Brass | ~450 (max use) | Outgasses Zn above ~450 C |
| Kaowool | >1000C | Flexible alumina/silica packing material |

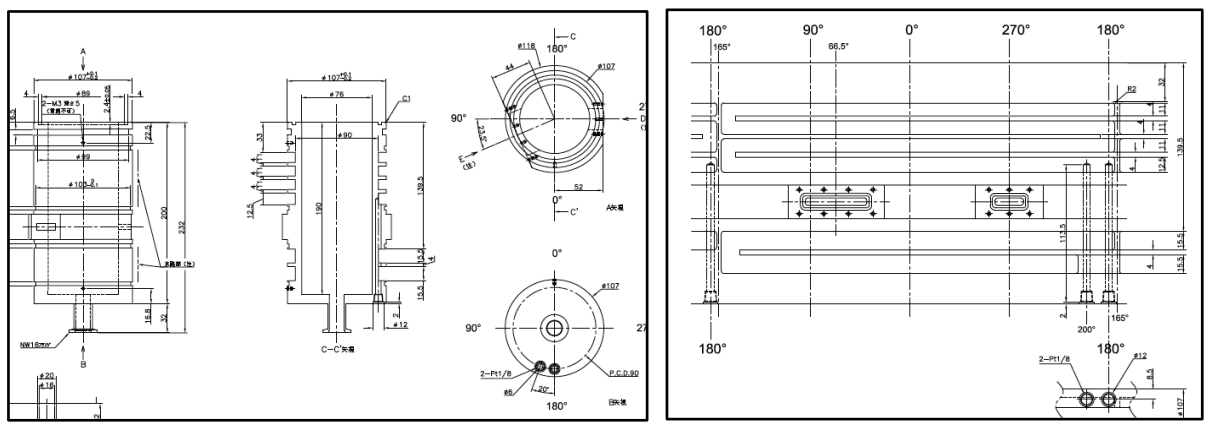

**Figure 1.** Water cooled vacuum vessel for the furnace machined out of a single piece of 5052 aluminum (machining by Satoseiki http://ssc-e.co.jp/)  The left panel shows several views of the vessel (a right cylinder 200 mm high with an outer diameter of 107 mm and an inner diameter of 76mm) while the right panel shows a planar projection of the cooling channels and x-ray windows locations.  The water channels are sealed using two outer sleeves that seal (tightly) with 4 o-rings – one at the top and bottom of each sleeve.  See also figure 6 for a picture of the chamber in use.

## 4. Chamber Design

The main part of the furnace chamber was fabricated from a single piece of aluminum (see **figure 1**).  It has intrinsic water channels that are fed from a chiller providing up to 2kW of cooling at room temperature.  The inner sample volume is 76 mm in diameter, while the outer extent of the main body, including windows, is within a diameter of 130mm (see also **figure 6**). The bottom of the chamber has an NW16 port used for pumping and/or He back filling, and effort was made to preserve a reasonable diameter (NW16 or larger) path to the vacuum pump to improve vacuum quality:  a critical goal for the furnace design, especially after issues with a previous setup, was robust vacuum performance.   The main sample unit is hung from a flange (see figure 5) that has 4 current feedthroughs (21032-01-W from MDC, https://www.mdcvacuum.com/, with 3.9 mm diameter copper conductors rated to 100A) welded in place (only two were used for the experiments discussed here), two dual thermocouple feedthroughs (part number 9313006 from MDC) on NW16 flanges and a pressure gauge on a ¼" Swagelok connector.  This unit performed well (after modification of the thermocouple feedthroughs discussed in section 11).

## 5. X-Ray Windows

The outer x-ray windows for the chamber are polished single-crystal sapphire, either 0.08 (incident) or 0.2 (outgoing) mm thick from Sartonworks (http://www.sartonworks.com/ ).  These are epoxied onto aluminum frames that are attached to the chamber using o-ring seals

(see **figure 2**, and also figure 1). The (1/e) attenuation length in sapphire is ~1.5 mm at 21.7 keV and ~2.5 mm at 25.7 keV, so 0.28 mm of sapphire will transmit 84% or 89% (at 21.7 and 25.7 keV, respectively) of the radiation. The use of the epoxy was regarded as speculative for high temperatures – but seems to work well. The windows are clear so they do not absorb much radiated heat from the interior of the furnace: even when the furnace is operated near 1600C they remain relatively cool so one can easily touch them with bare fingers (in contrast, the heat radiated through the window is large and will eventually burn a finger placed near the window). In operation, a layer of Al foil (12.5 um thick) is placed downstream of the outgoing window to insure that the thermal radiation exiting the chamber through the window does not heat the receiving Soller slit assembly (see figure 6). Also, at higher temperature we usually place a fan blowing on the window area to help keep the foil (and window) cool.

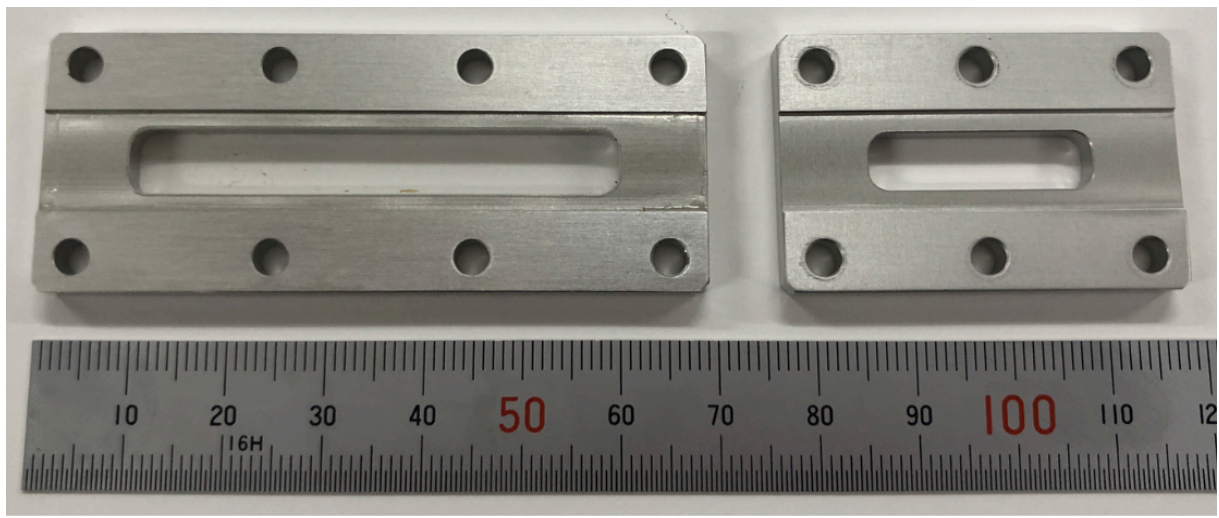

**Figure 2**: X-ray windows of polished sapphire epoxied onto aluminum frames. Scale is in mm. Left widow is the downstream window for scattered radiation and the right window is the upstream one for the incident beam. See also figures 1 and 6 showing the window placement on the chamber.

## 6. Heating Elements

After difficulties with short lifetimes of metal wire (W or Mo) heating elements, an alternative approach was desired. Options considered included carbon or carbon composite heaters as are compatible with running in vacuum or high-purity inert gas atmospheres, or $MoSi_2$ based heaters that are compatible with oxygen containing atmospheres. Notably, while robust in an oxygen-containing atmosphere, reportedly, $MoSi_2$ heaters can degrade after extended use in vacuum. Therefore resistive carbon-based heaters were chosen. Graphite heaters appeared to have an advantage in that they could easily be fabricated in three-dimensional shapes tailored to specific applications. However, most designs tended to have rather low resistances (<<1 ohm) so would need large current to achieve high power. It was also not clear how to attach the current leads so they would neither damage the graphite heater nor melt during use. Meanwhile, carbon composite heaters, essentially cut out of rigid sheet material, seemed to allow a simpler design, with modularly replaceable heater elements that were, in principle, stable to >2500C. Also, while significantly dependent on detailed design, it appeared a heater with a few (~2-5) ohms total resistance would be possible, relaxing the requirements on the current supply, and the current feedthrough. Carbon composite heating elements were initially purchased from Thermic Edge (https://www.thermic-edge.com/heaters/) and then later from Thermocera (http://www.thermocera.com/). One heater plate is shown in **figure 3**: the lower end of the heater, with the narrower, higher resistance, path, faces the Mo sample sleeve (see below) with one heater placed on each side of the sleeve (see figure 4). The nominal resistivity of the material is ~0.0016 Ohm-cm at 1000C. The measured resistance of two heater plates (two plates of the type shown in figure 3) used in series was ~4 ohms at room temperature dropping to ~2 ohms at 1600C sample temperature – with most of the resistivity change occurring below 400C.

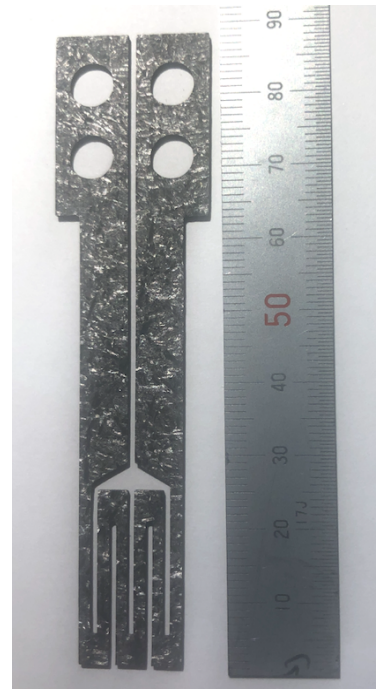

**Figure 3**. Carbon composite heating element. Scale in mm. Plate thickness is 1.2(1) mm.

## 7. Thermocouples & Thermocouple Feedthroughs

Type C thermocouples of W/Re were used (W/5%Re – W/26%Re) as they can be operated to temperatures in excess of 2200C. These were either bought pre-assembled (e.g. from Omega), or, more typically, fabricated directly from 0.25 mm diameter wire bought from Ishikawa-Sangyo (http://www.ishikawa-sangyo.co.jp/) that was twisted and spot welded. When needed, tubular ceramic insulators were used for the thermocouple wire (see figures 4 and 5). Thermocouple feedthroughs (part number 9313006 from MDC, https://www.mdcvacuum.com/) required some modifications (see section 11) but then performed well. Usually 2 thermocouples are used at one time, placed on opposite sides of the Mo sleeve just outside the sample: one of these is used for temperature feedback and the other is used for additional monitoring. Sometimes an additional thermocouple is also mounted on the sample cage to monitor its temperature.

## 8. Temperature Control and Power Supplies

We use Yokogawa UT35A temperature controllers (https://www.yokogawa.com/solutions/products-platforms/control-system/controllers-indicators/temperature-controllers/). These were chosen because they were ethernet and socket I/O compatible, and had the required range of functionality and compatibility with type C thermocouples. They performed well after some initial time investment to learn the format of the ethernet commands. A notable positive point was a convenient limit on the power used for auto-tuning, making them easy to auto-tune even at high temperatures. The output from the controller (0-20 mA over a 500 ohm resistor) was used to control the ITECH 6523D (0-200V, 0-60A, 3 kW, https://www.itechate.com/en/index_128.aspx ) DC power supply via its analogue (0-10V) input. DC supplies were chosen to reduce noise: another chamber with AC power supplies (and grounding through the chamber body) was known to occasionally introduce noise in the IXS detectors, while there has been no evidence of noise with the DC power supplies used here.

## 9. Sample/Heater Cage Design

The heater cage is shown in **figure 4**. A cylindrical tube of beryllium copper is used as the main outer support. The inner piece is a hollow rectangular BN block which is open at the top and bottom and has holes for the cylindrical molybdenum sample sleeve (see figure 4). The Mo sample sleeve design is described in [14]. Two carbon composite heater plates are supported from above and are located on either side of the sample facing the sleeve. Outside the BN block, two pieces of alumina are used as insulators, with 50 um thick molybdenum foils placed between the BN and the alumina to act as a reflector (one such Mo foil can be seen in the left panel of figure 4).

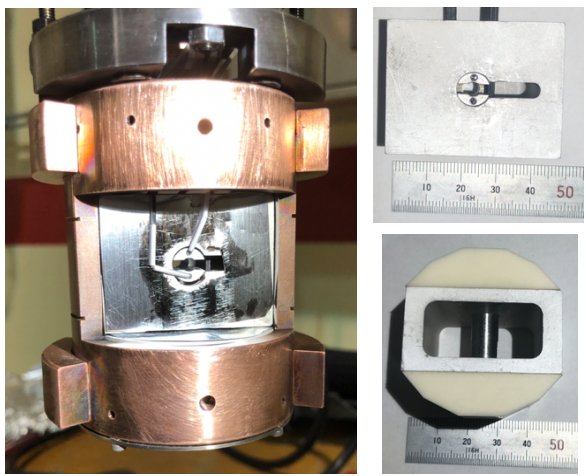

**Figure 4.** Sample cage and support. The left panel shows the entire cage including the outer beryllium-copper cylinder and the Mo sample sleeve with two thermocouple leads in ceramic insulators in the center. The upper right panel shows the inner BN sample box, the Mo sample sleeve (with holes for thermocouples, but no thermocouples) and the two heater plates incident from the top. Note the slot that allows measurement to larger, up to 55 degree, scattering angles. The right lower panel shows the white BN sample box viewed from above, with two pieces of alumina, one on each side. The Mo sample sleeve can be seen inside the BN box. The heater elements, not shown in the lower figure, would extend into the box, one on each side of the sample sleeve.

The sample space is covered above and below by disks of BN (inner layer) which are then covered by disks of alumina (BN and Alumina parts were fabricated by Krosaki, https://www.krosaki.co.jp/). The upper disks have slots for the heater elements and holes for the thermocouples.

Cooling is provided though 4 copper "fins" which, with thermal expansion, contact the water-cooled sides of the chamber (the 4 fins are visible in figures 4 and 5). Taking the coefficient of thermal expansion of the copper beryllium to be about 17 ppm/K and the outer diameter of 75 mm, one expects an expansion of about 130 microns in diameter per 100 C of temperature increase of the cylinder assembly. In fact, copper fins were carefully thinned by hand so they are a tight fit against the cooled chamber walls – with ~50 microns of clearance on each side.

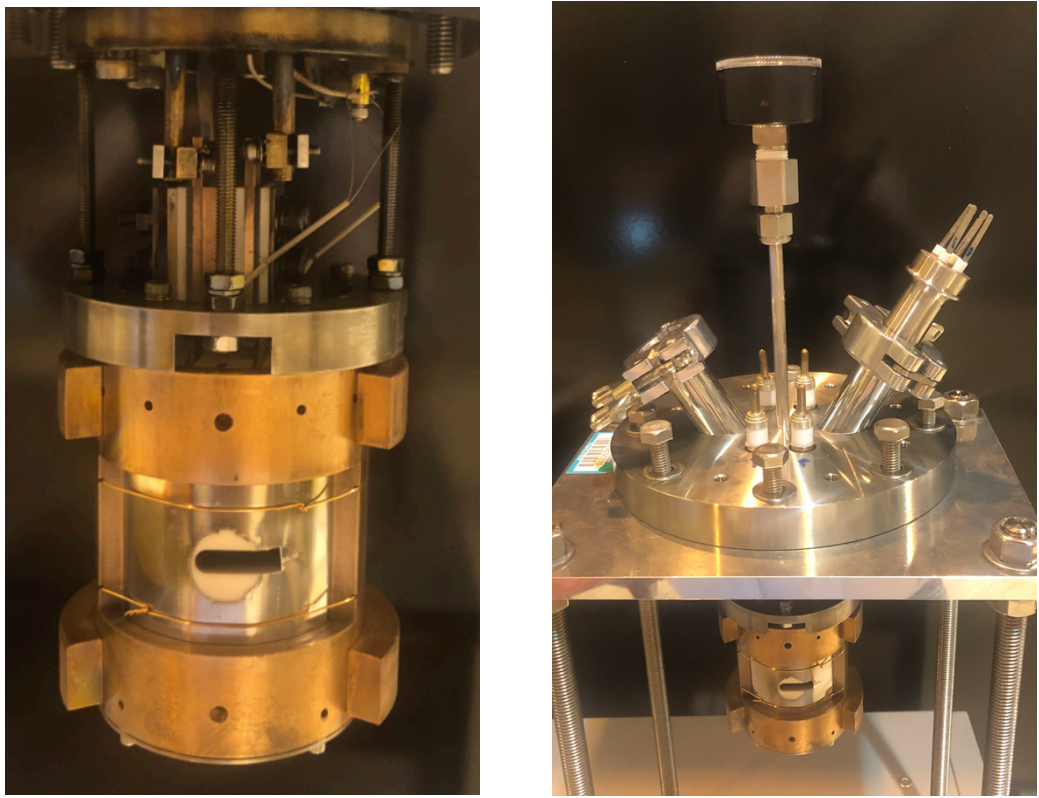

**Figure 5.** Sample cage mounted on the flange ready for insertion in the vacuum chamber. The left panel shows just the cage and leads while the right panel shows also the upper part of the flange including 4 current feedthroughs, one feedthrough for a pair of thermocouples, and the pressure gauge (see text for discussion).

In operation, one finds that with the sample position at 1600C, the temperature of the copper cylinder is about ~170C, so this cooling works well.

The assembled heater cage is shown in **figure 5** just before insertion into the water-cooled vacuum chamber of fig. 1. The horizontal slot for the scattered x-rays is visible in the center, partially covered by a 50 micron thick Mo foil over the alumina insulator. The foil is held in place on the cage by the two copper wires. The top of the heater assembly is visible above the cage, with the two cylindrical current leads visible at the very top of the figure. The tops of the two thermocouples (in ceramic insulators) are also visible at the upper right of the cylinder. Threaded rods are used to adjust the position (height) of the cylinder in the chamber to match that of the chamber windows.

## 10. Performance.

The performance of the system has been relatively robust. **Figure 6** shows the system mounted in the spectrometer. Operating in vacuum, one set of heaters has been used for multiple runs at temperatures of >~1500 C at the sample: the same heaters and thermocouples now have been used for a total of >18 days in these operating conditions (8 cycles from room temperature to >1500C and back) and continue to perform well. The pressure of the system measured at the pump cart (a system with a small, 50 l/s, turbo pump) typically is $<5 \times 10^{-5}$ Pa during the runs at 1500C, indicating the vacuum level is fairly good. The two thermocouples placed in the Mo sample sleeve read the same within about 25 C (worst case) near 1500C, with a corresponding estimated thermal gradient over the illuminated sample volume of < 1K  We typically calibrate the sample temperature by observing the melting point of the sample (e.g. pure iron melts at 1538C) using the change in the x-ray transmission. The transmission change is easy to observe on the first melting of a new sample. It is more difficult to observe when re-melting a sample that has already been melted (transmission changes on re-melting are sometimes only ~0.2%) but usually can be observed using high-rate detectors. Doing so shows that the sample temperature is typically about 50C lower than the average thermocouple temperature. This difference is reasonable given that the design of the sample holder allows some additional cooling for the sample (part of the sapphire sample cell – where it is sealed - extends well out of the molybdenum sleeve so remains relatively cool). When the heaters are turned off (during cooling), the two thermocouples read the same to within about 5C, suggesting the thermocouples are well matched and the difference in temperature observed at higher temperature is due to a real thermal gradient in the molybdenum sleeve. However, as this difference has not been worse than 25C (and in fact is often <10 C) in our measurement conditions, this is not considered serious. It probably originates from a slight difference in the performance of the two heaters or a slight asymmetry in their position relative to the sleeve.

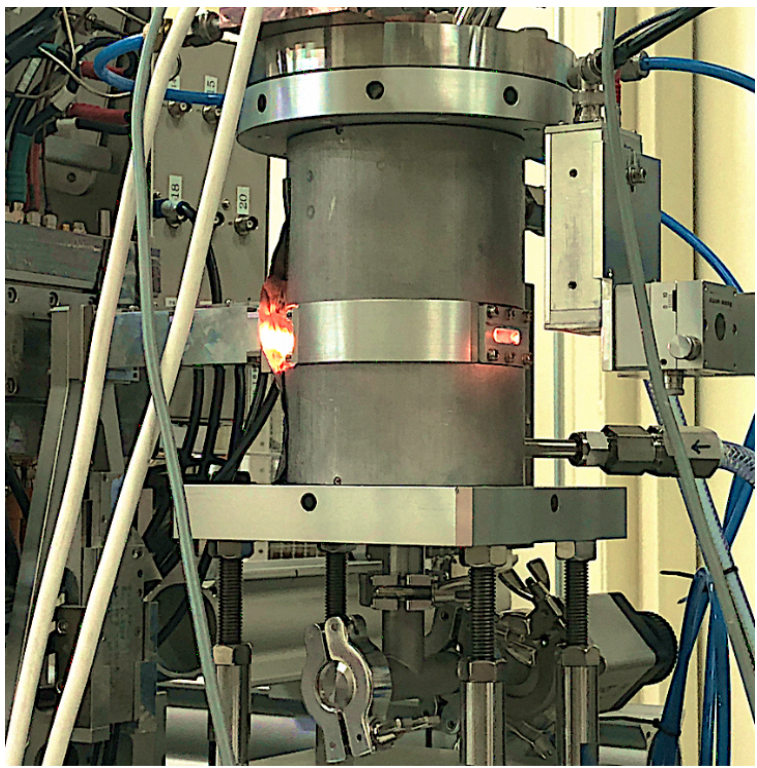

**Figure 6.** Furnace in use for IXS at BL43. The x-ray beam is incident from the right. Thermal radiation from the heater is visible coming through the windows. A piece of Al foil on the downstream side acts as a reflector to keep from heating the receiving Soller slit assembly. A cleanup slit and flux monitor for the incident beam are visible on the right side of the figure.

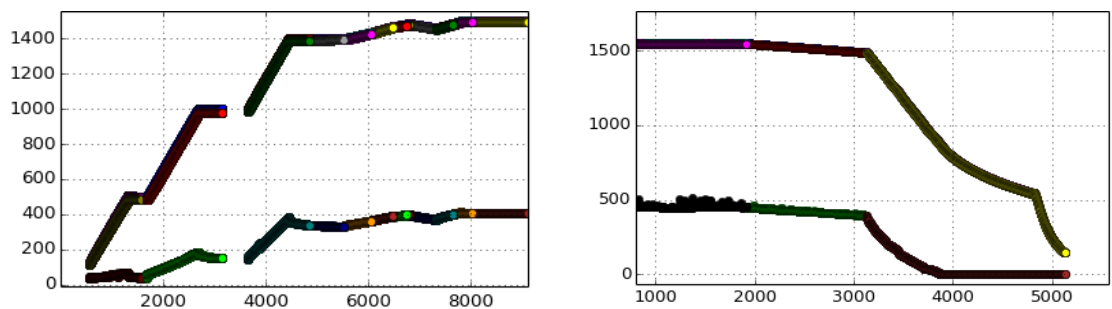

**Figure 7.** Heating and cooling of the furnace. The horizontal scale is time in seconds (relative to arbitrary zero times for each plot). The upper trace is the temperature of the two thermocouples in C (the two traces overlap) while the lower trace shows the heating power from the DC supply in watts.

Typical temperature and power curves are shown in **figure 7**. The system is usually first heated to ~400 C using a constant (but manually adjusted) voltage on the power supply. Then the system is ramped up to ~100C below the sample melting point at a rate of ~ 20C/minute. The heating is interrupted every 300 to 500 C to re-tune the system feedback parameters. Then the sample is usually slowly heated through the melting point (2 to 5 C/min) while observing the transmission. The melting point, as is usually observed by a sharp change in transmission, then becomes the reference for working at higher temperature. Cool-down usually occurs at 25 to 40 C/minute to ~600C. At that point the heaters usually have near zero current and a small amount of high purity He is introduced in the chamber to speed cooling. The control of the chamber is fairly easy above ~700C, but, below that, limited cooling for the system in vacuum limits the response time. (If temperatures <~700C are needed, operation in a low pressure He atmosphere is probably straightforward.)

The operation of the system is mostly automatic. It is usually monitored during the initial heating phase, and during the determination of the sample melting point, but then the controller is set at the desired operating point and the furnace control is not touched over typical run times of >24 hours. During that period the temperature at the control sensor usually remains very stable (within +- 0.3 K of the set point) while the drift of the second sensor varies : in a good run the full range of this drift is <2K, and, in the worst run to date, the full range was < 10K. Sometimes we also see jumps in both sensors (1 to 2 K in magnitude), that are quickly corrected by the control system. Thus we take the real sample temperature to be stable to +- 5K, or better, over ~24 hour time scales.

An IXS spectrum taken with this furnace is shown in **figure 8** from a sample of liquid iron (~0.1 mm thick) at T~1560C (about 22 degrees above the melting point) in a Hiroshima style sapphire cell [14] and measured using a Soller slit and specially designed analyzer masks [15]. This is an interesting sample for discussion about the nature and interpretation of the dynamics of liquids [2,3,5,6] and from the viewpoint of being a starting point to understand geological issues (e.g. [4]). The red points show the background (scaled by the sample transmission) and the blue points the measured scattering from the sample – the quasi-elastic and phonon contributions from the liquid iron are clearly visible (measurement at 25.7 keV with 0.75 meV energy resolution).

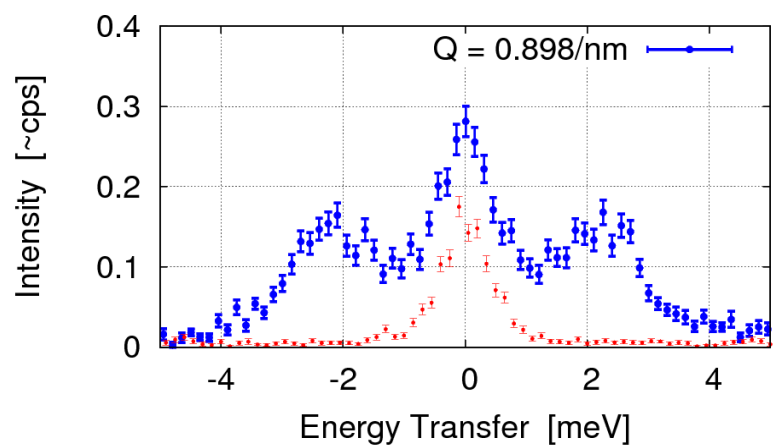

**Figure 8.** Measured spectrum of liquid iron at T=1560C and 0.898 nm$^{-1}$ momentum transfer using the furnace.

**11. Comments for Future Research and Development**

The furnace has performed well within the boundaries described above. However, through the process of constructing the furnace and making it work, some experience was gained that may be useful to others considering such a system in the future. This includes issues with thermocouple feedthroughs and also some issues related to going to higher temperature.

The thermocouple feedthroughs caused some difficulties. While their vacuum performance was good, the feedthroughs are "compensated" in that the conductors are not W/Re material, but something that is designed to be compatible over a limited temperature range (this was not so obvious from the description on MDC's website). These compensated wires extended far (~15 cm) beyond the vacuum interface. While the long extent facilitates the physical connections to the thermocouple leads, it also forced the connection points to be far enough inside the chamber that they viewed the furnace heating elements, at least partially. Thus, in initial work, the interface to the compensated wire was probably heated above the nominal operating limit and initial test runs with the chamber suffered from very poor and non-reproducible thermal control. After this issue was recognized, the long interior compensated leads were cut very close to the vacuum feedthrough and replaced by proper type C (W/Re) wires that were spot welded to the feedthrough. The interface between the compensated wire and the proper wire was then much further from (and out of view of) the heating elements. After doing this, the temperature control and reproducibility improved to the level discussed above.

With respect to attaining higher temperature, we note that during our first online usage of the chamber, we attempted to go to 1900C at the sample, and this led to the heaters failing in about 4 hours (they were replaced), even though they had been very stable at ~1600C. This has not been explored further, as, so far, the samples of interest have melted below 1600C. Also, usling a slightly different design (with 3 heater plates), we melted the molybdenum sample sleeve ($T_M$=2623C) during off-line tests (this over-power event was probably the result of using the compensated thermocouple feedthrough in its original configuration). Thus it appears the carbon composite heating elements may indeed function over 2000C, but achieving temperatures higher than the 1600C used here would require some additional work.

We note that at higher sample temperatures the current feedthroughs become warm. The flange is cooled, but, apparently, more cooling at the feedthrough might be good. The source of the heat is not the direct resistive heating by the current flowing through the feedthroughs, but the heat conducted from the heater region inside the chamber back to the feedthrough: a low-resistance current path is usually also a low-resistance heat path. The temperature of the feedthroughs was reduced by putting a small section of W wire (0.6mm diameter x ~10mm long) between the feedthrough and the main heater body, increasing the thermal isolation of the feedthrough from the heater. At a sample temperature of ~1600C the feedthroughs are then warm to touch. This heating might become an issue for operation at extended times at temperatures >1600C.

## 12. Concluding Comments

This paper has described the design and operation of a furnace for inelastic x-ray scattering that operates robustly up to temperatures of 1600C. A pair of carbon composite heating elements have, as of the time of the writing of this paper, survived >8 heating cycles (RT to >1500C) with an integrated time at T>1500C of more than 18 days. The use of Hiroshima type single crystal sapphire cells [14] along with single crystal windows and a careful IXS setup, allows measurements to be conducted even to sub-inverse-nm momentum transfers at high temperature.

## 13. Acknowledgments